\documentclass[11pt,letterpaper]{article}
\usepackage[final]{graphics}
\usepackage{graphicx}
\usepackage{makeidx}
\usepackage{authblk} 
\usepackage{epstopdf}
\usepackage{listings}
\usepackage{pstricks,tikz}
\usepackage[absolute,overlay]{textpos}

\usepackage{amsmath, amssymb}
\usepackage{wrapfig}
\usepackage{color,bm}
\usepackage{natbib}
\usepackage[applemac]{inputenc}
\usepackage[portrait]{geometry}
 \usepackage{setspace}

\begin{document}

\title{Interface Retaining Coarsening of Multiphase Flows}

\author{Xianyang Chen, Jiacai Lu and Gretar Tryggvason}
\affil{Johns Hopkins University, MD,  USA}

\date{\today}

\maketitle


\begin{abstract}
Multiphase flows are characterized by sharp moving interfaces, separating different fluids or phases. In many cases the dynamics of the interface determines the behavior of the flow. In a coarse, or reduced order model, it may therefore be important to retain a sharp interface for the resolved scales. Here, a  process to coarsen or filter fully resolved numerical solutions for incompressible multiphase flows while retaining a sharp interface is examined. The different phases are identified by an index function that takes different values in each  phase and is coarsened by solving a constant coefficient diffusion equation, while tracking the interface contour. Small flows scales of one phase, left behind when the interface is moved, are embedded in the other phase by solving another diffusion equation with a modified diffusion coefficient that is zero at the interface location to prevent diffusion across the interface, plus a pressure like equation to enforce incompressibility of the coarse velocity field. Examples of different levels of coarsening are shown. A simulation of a coarse model, where small scales are treated as a homogeneous mixture, results in a solution that is similar to the filtered fully resolved field for the early time Rayleigh-Taylor instability.
\end{abstract}

\section{Introduction} 
Multiphase flows, just like single phase flows, are usually unsteady, with a large range of spatial and temporal scales. Although the governing equations are known and it is in principle possible to numerically simulate the time dependent evolution, for most problems of practical interest the large range of scales makes doing so impractical. Experience also suggest it is unlikely to be necessary since in most cases the flows exhibit a fair degree of universality at the smallest scales. For single phase flows it is by now fairly well recognized that while it is  difficult to develop general closure laws for mathematical models of the fully stationary average flow, models where the unsteady motion of the large scales is simulated and models are only used to describe the average motion of the small scales, can lead to much improved predictions. Indeed, in many cases it is found that models for the small scales can be much less elaborate  than closure models for the average motion, since the range of scales that needs to be modeled is greatly reduced. This approach is often referred to as Large Eddy Simulations (LES). LES originally referred to  simulations where only very small scales were modeled, but it is now often used indiscriminately for any unsteady simulation incorporating models for unresolved scales.  For multiphase flows, where a sharp interface separating large regions of different phases, it is likely that a similar approach, where the unsteady motion of the large scales is tracked and small scales are modeled, is the most realistic way to develop accurate predictive strategies. 

Considerable effort has been put into developing LES like models for multiphase flows. \cite{SagautGermano2005}  discussed filtering flow fields near interfaces, noting that the filtering  should smooth the turbulence but preserve the discontinuity, and derived  consistency conditions for possible subgrid models. Filtered LES equations were derived in \cite{Labourasseetal2007}, who identified the unresolved terms, and showed results of an {\it a priori} test, where proposed closure terms are compared with filtered DNS results for two two-dimensional flows, one with small vortices interacting with a bubble and another for ``phase inversion'' where a light fluid at the bottom of the computational domain moves to the top. This study was extended to three-dimensional flows by \cite{Toutantetal2008} who proposed closure relations for some of the subgrid terms and examined the interaction of small scale vortices with a single deformable bubble. Further studies of the fully three-dimensional phase inversion problem was done by \cite{Vincentetal2008} who also derived the averaged LES equations and examined the magnitude of the various subgrid terms.   
The jump conditions for interfaces in filtered fields were revisited by \cite{Toutantetal2009a} and \cite{Toutantetal2009b} who in addition to filtering the turbulent flow simplified the interface, and proposed subgrid models for some of the unresolved terms.
Model equations were also derived by  \cite{LiovicLakehal2007} who used ``component-weighted volume averaging'' to filter the equations and develop models for unresolved subgrid terms  near deforming interfaces, based on the distance to the interface.  \cite{Aniszewskietal2012} used approximate deconvolution to model the surface tension tensor for the filtered equations and \cite{WaclawczykOberlack2011} discussed modeling in the context of ensemble-averaged fields.
A very different approach to subgrid  modeling was introduced by  \cite{HermannGorokhovski2008b,HermannGorokhovski2009} and \cite{Herrmann2013} who used a dual scale strategy where the flow field is coarsened but the full interface is retained on a finer grid.  The filtered equations are similar to those in \cite{Toutantetal2009a}  but the subgrid terms for surface tension can be computed directly for the well resolved interface. A subgrid model is, hoowever, need  for the advection of the interface on the finer grid.
Recent development of closure terms for the model equations derived in \cite{Labourasseetal2007}, focusing on atomization, can be found in \cite{KetterlKlein2018,Ketterletal2019a} and \cite{Ketterletal2019b}. 
Studies of the use of deconvolution models for a two-dimensional phase inversion are described in \cite{Gouenardetal2019} and for attempts to deal with non-isothermal flow see \cite{Saeedipouretal2019}. 
A large number of authors have developed numerical models where small bubbles and drops are represented as Lagrangian point particles while interfaces are tracked using any of the many numerical method for flows with sharp interfaces (\cite{Tomaretal2010,Maetal2017,Hsiaoetal2017,Zuzioetal2018}, for example). 
Often the point particles are formed by breaking off a small fluid blob from an interface, usually in an ad-hoc manner. The interface separating two fluids consists, however, not just of relatively ``resolvable'' part and fluid particles, but usually has waves, splashes, crowns, finger and folds---some of which may lead to bubbles or drops and others that do not---but that exists at scales too small to be resolved in simulations of real systems. Another effort to retain large scale structure but model small scales can be found in \cite{Nykterietal2020} who use a hybrid method where a sharp interface method is used for some part of the computational domain, and a $\Sigma-Y$  two-fluid model, introduced by \cite{Navarro-Martinez2014},  is used for the unresolved parts, with a dynamic switching between the different approaches. 
Recent  discussions of the status of LES modeling for multiphase flows can be found in  \cite{Vincentetal2018, Lakehal2018, Mukundanetal2020} and \cite{Nykterietal2020}.

As \cite{Toutantetal2009a}  point out, coarse models are significantly different from single phase LES models because of the presence of the index function and a few author have sought to make that clear by suggesting different names. Thus, \cite{Toutantetal2009a} talk about the Interfaces and Sub-grid Scales (ISS) approach and \cite{LakehalLabois2011} about Large-Eddy \& Interface Simulation (LEIS) models. Later papers seem, however, to have settled on referring to methods based on filtering as ``LES for multiphase flows.'' Here we will refer to the coarse field and talk about coarse flow models.

For single phase flows, coarsening the flow field by filtering to eliminate high wave number content is a well-established procedure, although accounting for the effect of the small scales by a closure model is still ongoing research. 
The main challenge for multiphase flow is that filters for single phase flows smooth everything, including the phase boundaries. Since the interface separating the fluids is often a dominant feature of the flow, it seems that a better approach would be to coarsen the flow in a way that retains the interface, at least in many parts of the flow. Interface retaining coarsening is a common issue in many areas such as cartography where a sharp shoreline, although simplified, is retained on large-scale maps, and in image processing where it is found that retaining interfaces and steep gradients gives ``better'' coarsened images (\cite{Paletal2015}).
Apart from simulations where small droplets are broken off the liquid as Lagrangian drops in simulations of atomization (\cite{Tomaretal2010} and other references cited above)  and the work of \cite{Nykterietal2020}  where small scales are represented by a two-fluid model, we are only aware of two efforts to retain sharp interfaces in modeling of multipahse flows. In \cite{Toutantetal2009a} the index field identifying the different phases is first  smoothed by applying a Gaussian filter and then the interface is reconstructed by identifying the contour  that originally coincided with the interface. Recognizing that the interface boundary condition can change,  \cite{Toutantetal2009b} used asymptotic expansion to derive new boundary conditions for flows with bubbles, for situations where the filter size is much smaller than the bubbles, justified by assuming that although the bubble surface is disturbed by the turbulence, the Kolmogorov length scale is much smaller than the bubble diameter.
The other discussion of interface retaining coarsening is a very short section in \cite{Lakehal2018} where an ``All-Regime Multi-fluid model'' is introduced. Although preliminary results are shown, little details and no reference are provided and no further development seems to have taken place. In all other studies of coarse models for multiphase flows, referenced above, the index function seems to have been filtered along with the rest of the flow field.

Development of closure models in single phase flows have traditionally relied on analytical models that are calibrated with results from direct numerical simulations where the unresolved terms can be computed exactly, since their form, in terms of the resolved and the filtered variables is known. Most authors developing models for multiphase flows have taken a similar approach. Recent developments in a number of areas, which we will collectively refer to as data-driven modeling, however, suggests a different approach. Several authors have, in particular, suggested building partial differential equations directly from data (\cite{RaissiKarniadakis2018,Longetal2018,Leeetal2020}). In this approach the ``learned'' equations ensure that the coarse field evolves correctly, and the fully resolved field plays no role except as the starting point for the coarsening. For fluid flows we have, of course, a fairly good idea what the overall structure of the equations are, so instead of having to learn the full equations, we should only have to learn how to modify them or add extra terms.  Machine learning can, of course, also be used to find closure terms from the fully resolved solution (see \cite{Maetal:PF:15,Maetal:IJMF:16}), but the possibility of obtaining the modifications directly from the coarse data offers possible new strategies. Indeed, one of the main consequences of approaching closure modeling from this perspective is that the coarsening can be done in many ways and we do not need an explicit connections between the extra terms and the fully resolved field. For a simple demonstration of this approach for 2D single phase turbulence, see \cite{Chenetal2021}.

Here we explore coarsening the phase distribution and the flow field where a sharp but simplified interface is retained. The index function is coarsened by solving a constant coefficient diffusion equation, while tracking the interface contour. Patches of one phase, left behind when the interface is moved, are embedded in the other phase by solving another diffusion equation with a modified diffusion coefficient that is zero at the interface location to prevent diffusion across the interface. When smoothing the the momentum, we  enforce incompressibility of the coarsened flow field by also including a pressure field. The small scales can be treated in many ways, including by the models referred to above, but here we only give one example, using a simple homogeneous mixture model with closure models determined by inspection. 

\section{Scope} 

We consider unsteady incompressible flow consisting of different fluids or phases governed by the Navier Stokes equations
\begin{equation}
\frac{\partial \rho {\bm u}}{ \partial t} + \nabla \cdot (\rho {\bm u} {\bm u} )=-\nabla p+\rho {\bm g} +\nabla \cdot \mu (\nabla {\bm u} +\nabla {\bm u}^T) + {\bm f}_\sigma \quad  \hbox{and}\quad \nabla \cdot {\bm u} =0.
\end{equation}
Here, $\rho$ is the density, $\mu$ is the viscosity, ${\bm u}$ is the velocity, $p$ is the pressure, $ {\bm g}$ is the gravity acceleration and $ {\bm f}_\sigma$ is the surface tension term. Both the density and the viscosity are different in the different fluids. Solving these equations accurately gives the fully resolved flow field at any given time and spatial location. 
The different phases are identified by an index or marker function $\chi$ and assuming, for simplicity, that only two phases are involved, we have:
\begin{equation}
\chi ({\bm x}) = \left\{ 
  \begin{array}{l l}
\text{0 in fluid 0}\\
\text{1 in fluid 1}.
  \end{array} \right.
\end{equation}
The various flow quantities, such as density, are then given by $\rho = \chi \rho_1 + (1-\chi) \rho_0$. 

\begin{figure}
\centering { \includegraphics[scale=0.37]{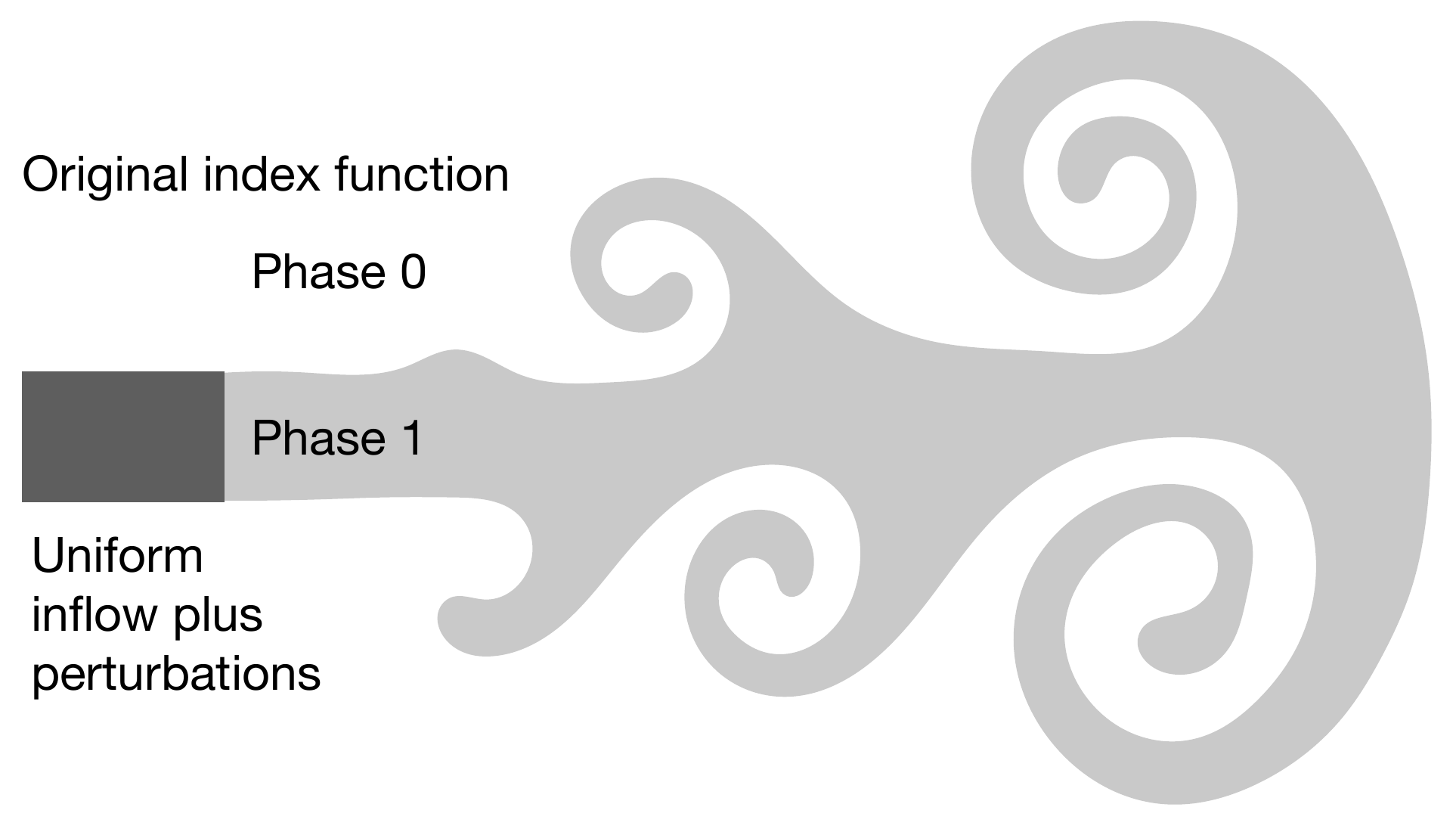}   \includegraphics[scale=0.37]{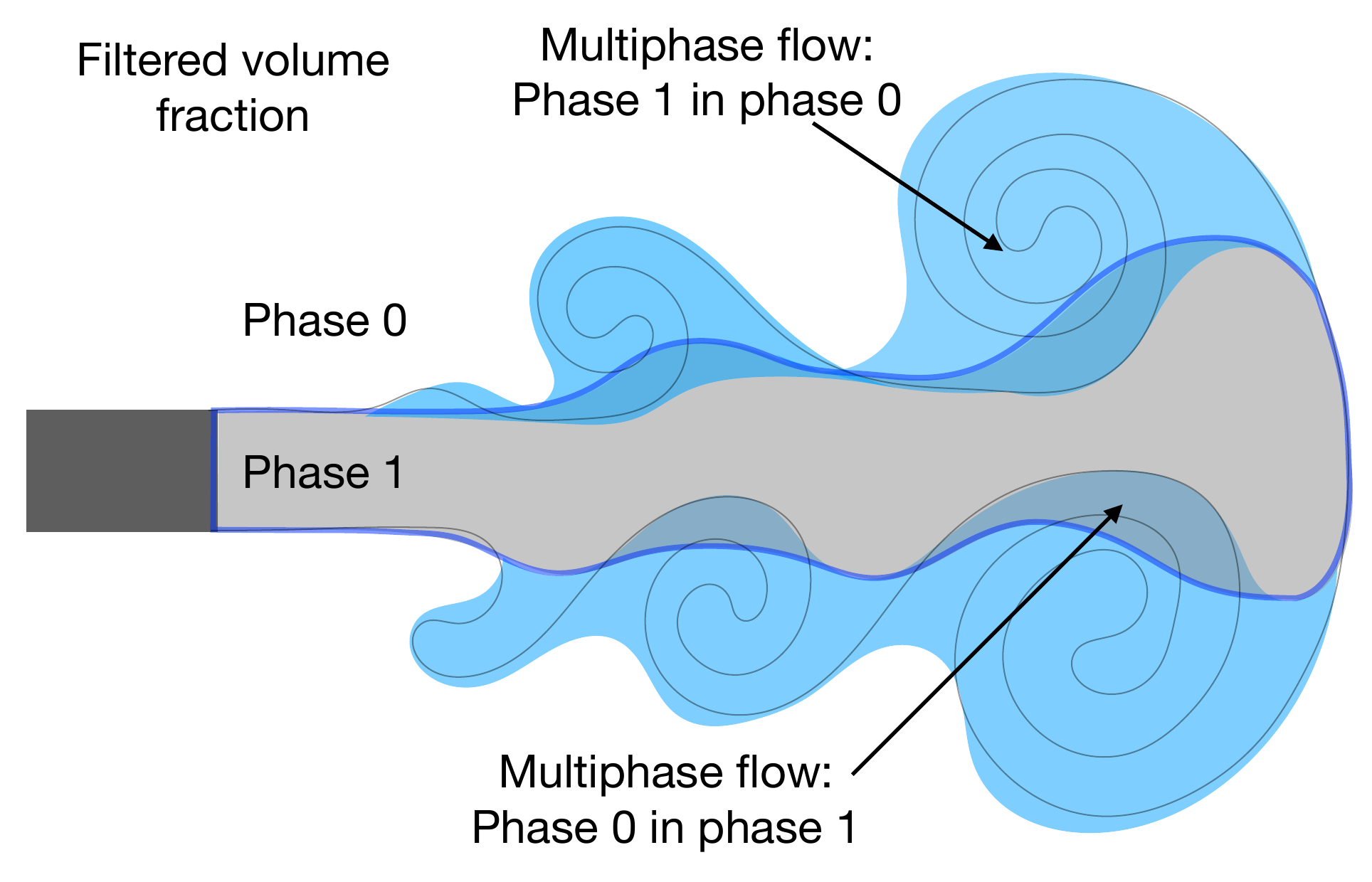} }
\caption{Aggressive filtering of an idealized starting jet. The original index function is shown in the left frame and the filtered index function is shown on the right. The large scale flow is represented by the gray area and the mixed zones by the blue regions. The original interface is shown by a thin solid line.}
\label{Filtering-1}
\end{figure}

The full solution generally contains a large range of scales and the purpose of a coarser model is to remove the smallest ones, yet account for their effects. For a single phase flow where filtering to remove the high wavenumber components is generally used, the coarse flow is simply a smoother version of the full solution. For multiphase flows, where a sharp interface separates the phases, one possibility is to smooth the index function along with the velocity field, thus removing the sharp interface. Here, however, we pursue a different strategy and retain a sharp interface, although its shape will generally be simplified. Thus, we seek a coarsening strategy that has the following  characteristics:
\begin{itemize}
	\item The large and the small scale phase distribution are separated by coarsening the index function. The interface is simplified but stays sharp, with some parts ``collapsing'' into points, lines or sheets of zero volume.
	\item The large and the small scale flow on either side of the interface are separated by coarsening.  In regions where the smoothing of the index function causes fluid to ``switch sides,'' the fluid is mixed with or embedded in the other fluid. 
	\item As the coarsening is reduced, the flow field  approaches the results given by direct numerical simulations (DNS).
\end{itemize}
Ideally, the coarsening  strategy should be general in that we can coarsen just a little bit as well as very aggressive.
Figure \ref{Filtering-1} shows the process schematically.  Here we assume a very aggressive filtering so that the coarse index function is much simpler than for the fully resolved flow. As we filter the index function, we also filter the flow variables, creating mixed zoned on either side of the interface, as shown on the right. Notice that while the densities and other material properties in the original fully resolved flow may be constant,  that is not generally true for the coarsened flow.

The coarsening is a precursor to the development or implementation of multiscale models to evolve the coarse flow field. 
The flow in the mixed zones can be modeled in a variety of ways, ranging from simple mixture models, drift flux model, full two-fluid models or by using Lagrangian point particles to account for bubbles in liquids and drops in gas. Away from interfaces, models for the flow field should become standard models for multiphase and/or turbulent flows but close to the interface models need to developed for the additional terms, using analysis and/or machine learning.

\section{Coarsening the Fully Resolved Field} 

We start by the observation that the solution of the linear diffusion equation 
\begin{equation}
\frac{\partial g (t,{\bf x})}{\partial t} =D  \nabla^2 g(t,{\bf x}),
\end{equation}
in an unbounded two-dimensional domain, is given by
\begin{equation}
g({\bf x},T)= \frac{1}{ 4 \pi D T}  \int_{Area} e^{- \frac{|| {\bf x} - {\bf x}' ||^2 }{ 4 D T}} g_o ( {\bf x}') da',
\end{equation}
at time $T$, where $g_o$ is the initial condition.
Thus, filtering a field by a Gaussian kernel is equivalent to evolving it by a diffusion equation for a time that is related to the length scale of the filter $\Delta$ by $ 4D T=  \Delta^2 /6 $. The reason we find it more convenient to coarsen the field by diffusion rather than filtering, should become clear in the rest of this section. 
The connection between filtering and diffusion for single phase turbulence has been pointed out by \cite{Johnson2020,Johnson2021},  and \cite{CapecelatroDesjardins2013} use diffusion to distribute the effect of small particles onto a fixed grid, as examples of other uses of diffusion instead of explicit filtering. We have taken advantage of the flexibility of doing smoothing by diffusion rather than filtering in \cite{Chenetal2021a}, where we used it to sharpen a smooth distribution.

\begin{figure}
\centering { \includegraphics[scale=0.6]{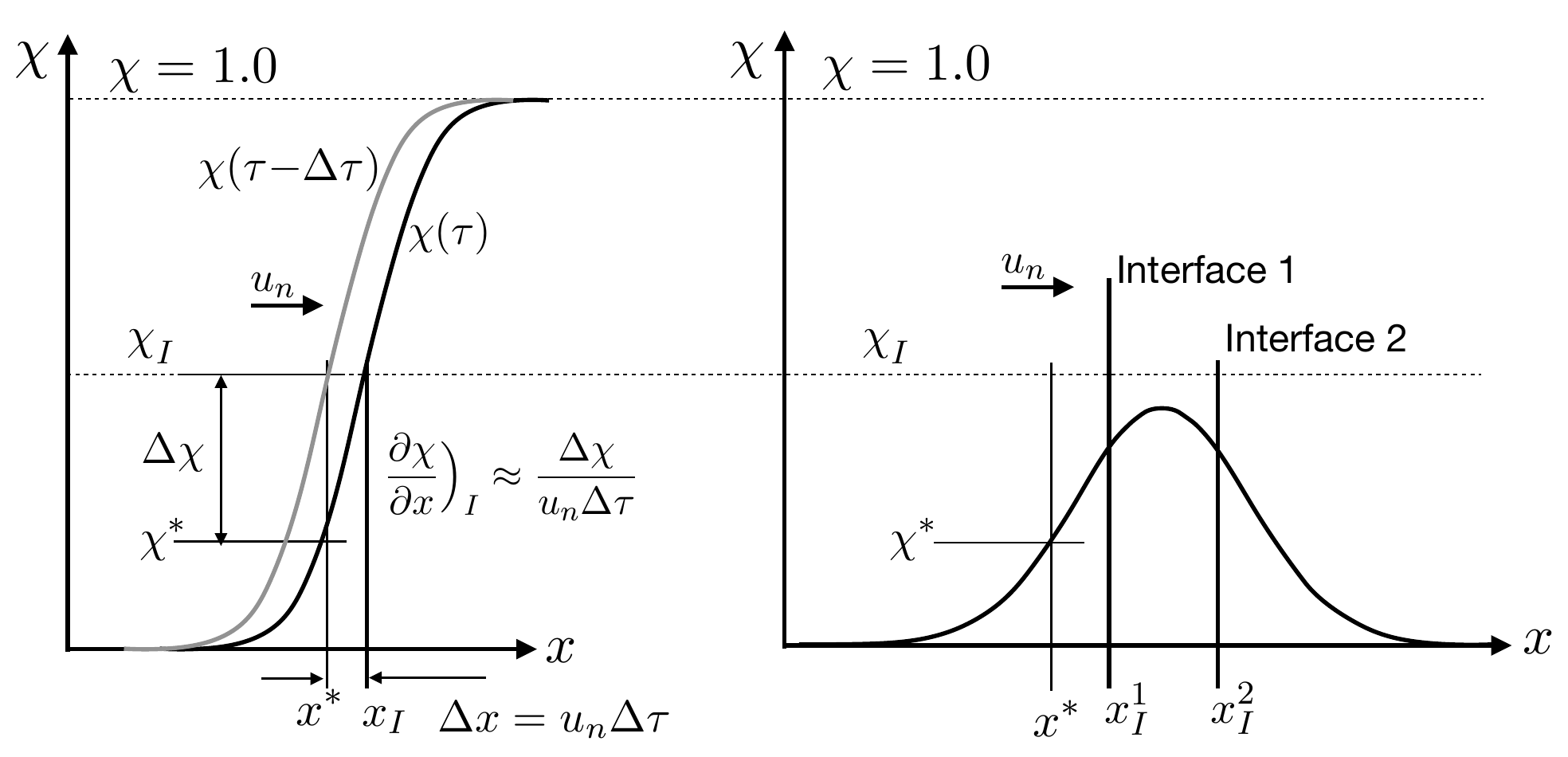} }
\caption{The motion of an interface following the interface contour of the index function. Left: the interface, where the index function has the value $\chi^*$, 
needs to catch up to $\chi_I$ and must move a distance  $\Delta x =u_n \Delta \tau$, computed from $\chi_I$ and the gradient.  Right: Once interfaces are close enough, there is no $\chi_I$ contour close by. }
\label{sketch1}
\end{figure}

\subsection{Smoothing the Index Function}

To simplify the interface we diffuse the index function by solving
\begin{equation}
\frac{\partial \chi }{ \partial \tau} =  \nabla^2 \chi,
\label{diffusion1}
\end{equation}
in ``pseudo-time'' $\tau$, until $\tau=T$, at every instance in ``real time'' when we want a smoothed solution. Notice that we set the diffusion coefficient to unity ($D=1$) since a different value simply rescales $\tau$.
To move the interface with the contour identifying the interface, $\chi = \chi_I$, we find the ``diffusion velocity," using that the value of the function at the interface does not change, so the material derivative of $\chi$ is zero:
\begin{equation}
\frac{D \chi}{ D \tau} =\frac{\partial \chi}{ \partial  \tau } + {\bm u} \cdot \nabla \chi =0.
\label{advection1}
\end{equation}
We only need the velocity in the normal direction to the interface, and a normal is defined by ${\bm n} =\nabla \chi / \vert \nabla \chi \vert$, so we write ${\bm u}= u_n {\bm n} = u_n ( \nabla \chi /\vert \nabla \chi  \vert ) $. Rearranging and dividing by $\vert \nabla \chi  \vert $ gives
\begin{equation}
\frac{1 }{ \vert \nabla \chi  \vert } \frac{\partial \chi}{ \partial  \tau} =-u_n \frac{ \nabla \chi \cdot \nabla \chi}{ \vert \nabla \chi  \vert^2} =-u_n,
\end{equation}
since ${\bm n} \cdot {\bm n}=1$. After the index field has been updated by taking one step in pseudo-time, the interface usually no longer coincides with the interface contour $\chi_I$ and for an interface point that is not exactly on the interface contour,  the value differs by $\Delta \chi = \chi_I - \chi^*$ from the interface value, where $\chi^*$ is the old value of the index function at the interface. See figure \ref{sketch1}. In a time step $\Delta \tau$ the interface thus needs to ``catch up,''  so we approximate $\partial \chi / \partial t \approx \Delta  \chi / \Delta  \tau$ and write
\begin{equation}
{\bm u}_I=u_n {\bm n} = -\frac{ (\chi_I - \chi^* )  }{ \vert \nabla \chi  \vert^2  \Delta  \tau }  \nabla \chi.
\label{interfacevelocity1}
\end{equation}
We interpolate $\chi$ from the grid to find $\Delta \chi $ and $\nabla \chi$, from which we can find $\vert \nabla \chi  \vert^2 = \nabla  \chi   \cdot  \nabla  \chi $. 
Each interface point is then moved by solving
\begin{equation}
 \frac{d {\bm x}_I }{ d \tau} = {\bm u}_I.
\end{equation}
\begin{figure}
\centering { \includegraphics[scale=0.265]{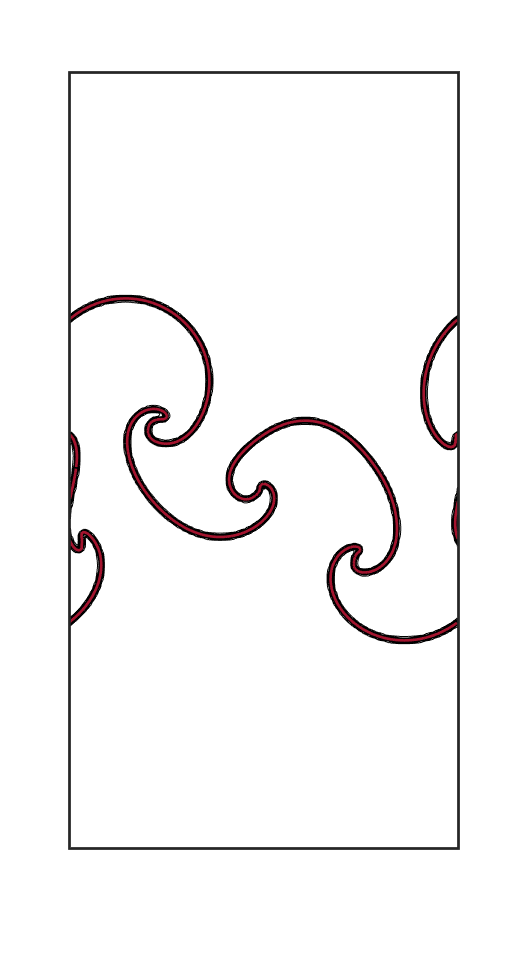}    \includegraphics[scale=0.265]{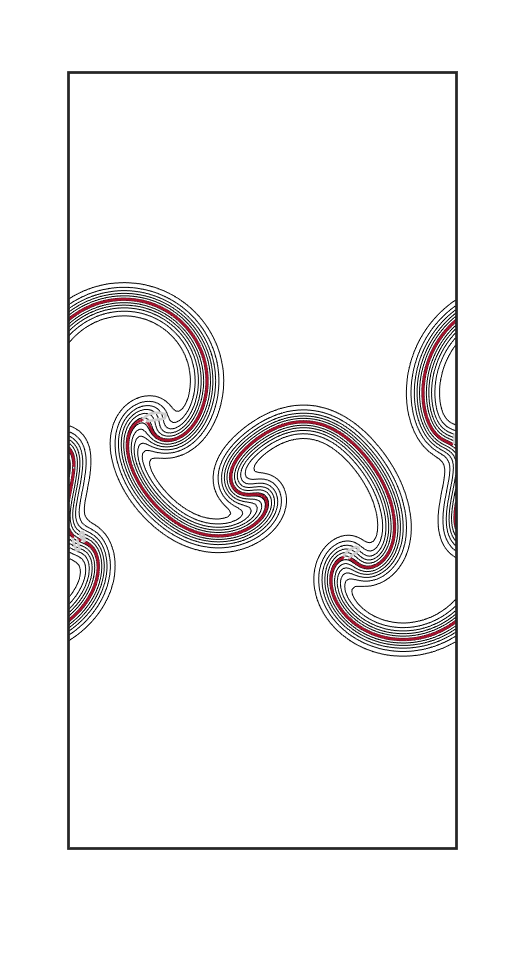} \includegraphics[scale=0.265]{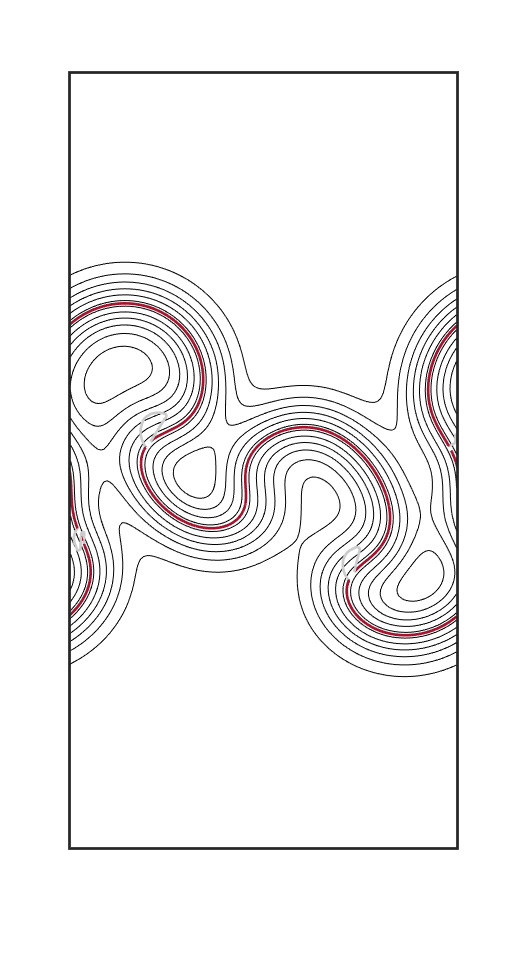}  \includegraphics[scale=0.265]{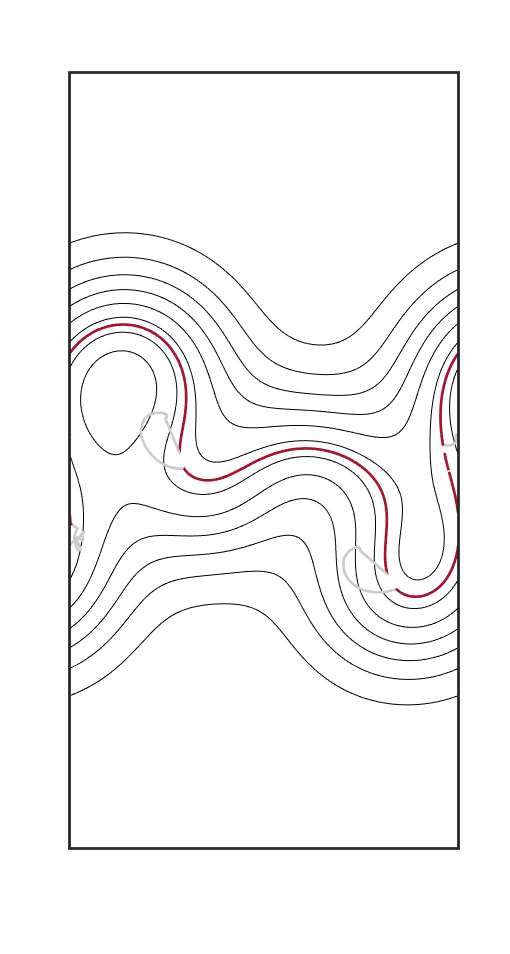}     }
\centering { \includegraphics[scale=0.265]{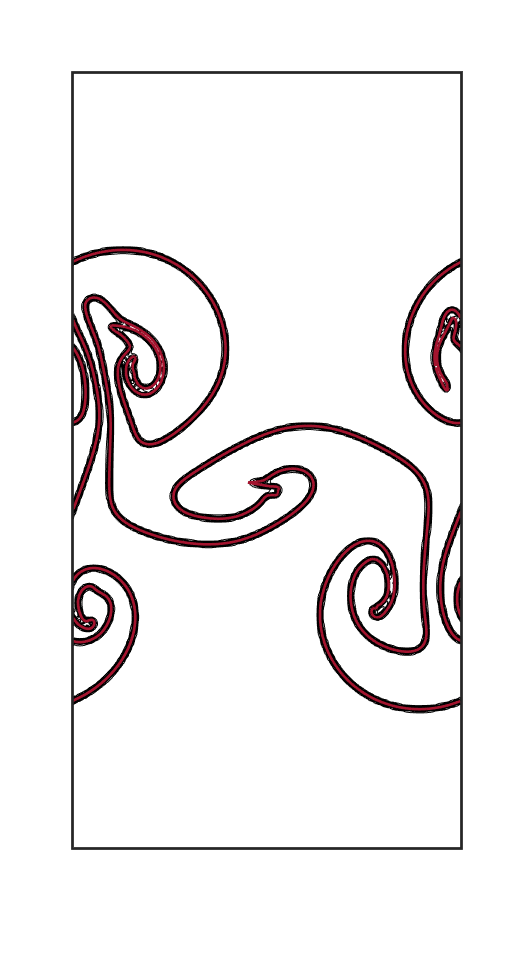}    \includegraphics[scale=0.265]{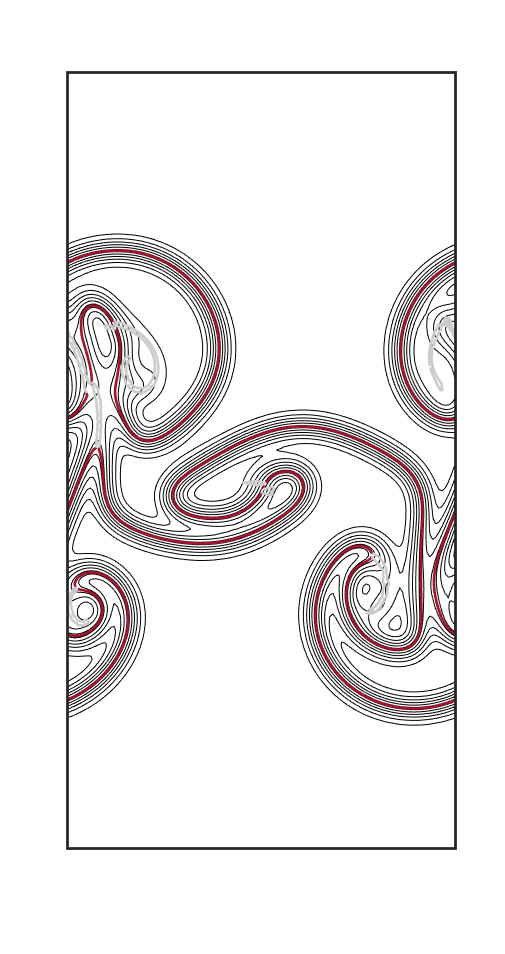} \includegraphics[scale=0.265]{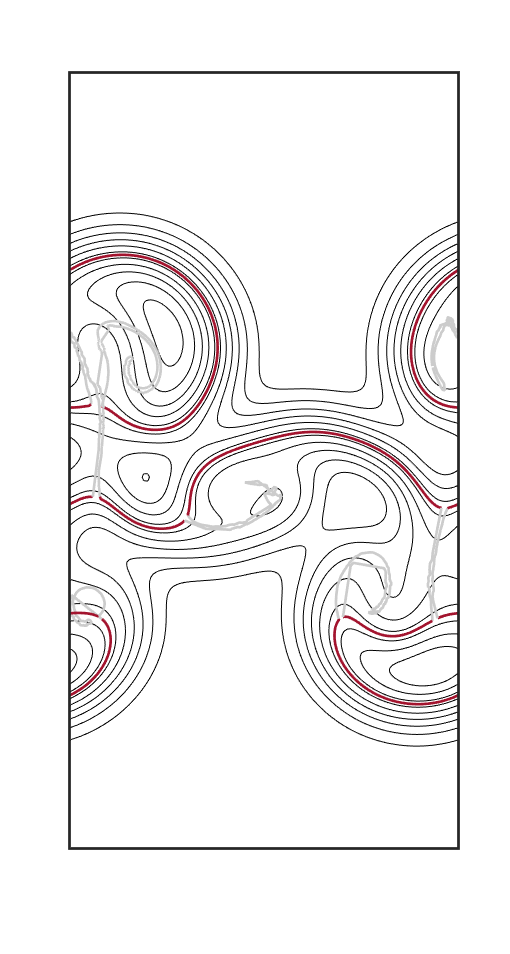}  \includegraphics[scale=0.265]{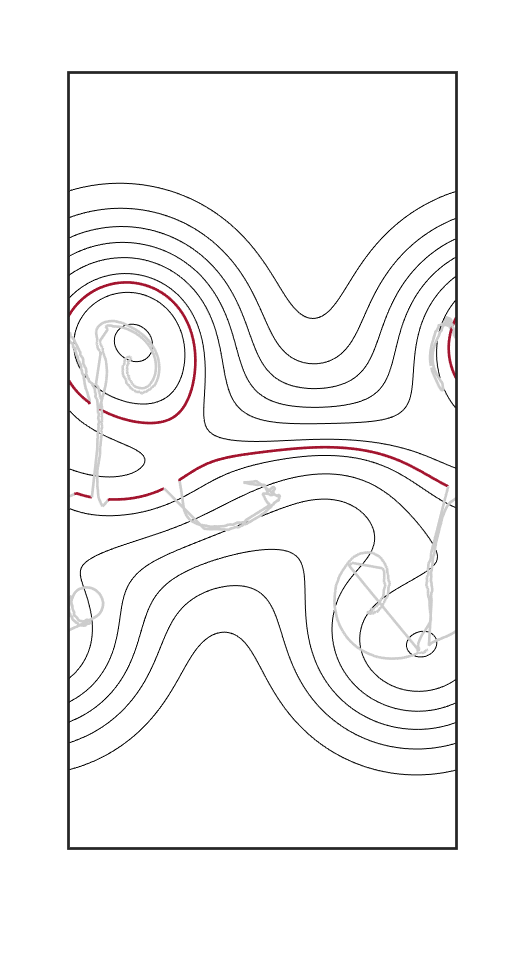}     }
\caption{The evolution of an interface simplified by diffusing the index function at times $t=0.175$ (top row) and $t=0.25$ (bottom row). In each row the results are shown for $\tau=0$ (original), $\tau=0.0004$ ($\Delta = 0.1$), $\tau=0.0026$ ($\Delta = 0.25$), and $\tau=0.0104$ ($\Delta = 0.5$). The interface is the red line and  contours of the index function are also shown. ``Collapsed'' interfaces are shown as gray lines.   }
\label{Figure-Interface}
\end{figure}
An interface moves  when the index function is smoothed by diffusion for two reasons: Curvature of  the interface and the presence of a close, parallel, interface. Solving a diffusion equation accounts for motions due  to both reasons, but whereas an interface will simply stop moving when it becomes straight, an interface moving toward another interface is likely to eventually cross since the extreme value of the index function will be smaller or larger than the interface value and there is therefore no nearby point where the value of $\chi$ is equal to the  interface value. In those cases it is incorrect to approximate the time derivative as we did above and we simply want the interface to stop moving. To check for this, we compute a normal to the interface and  examine the value of $\chi$ at a point slightly ahead of the interface. In most cases this value will be larger than the current value of $\chi$, if $\chi$ is smaller than the interface value, and smaller if $\chi$ is larger than the interface value, but if both values are  on the same side of the interface value, then it is likely that another interface is nearby and we put the interface velocity  to zero. We usually also flag such ``collapsed'' interfaces since often we may want to ignore them when modeling the mixed zone. 
We note that our procedure for smoothing the index function is essentially identical to the one used by \cite{Toutantetal2009a}, except that we follow the interface as the field is smoothed, instead of first smoothing and then restoring the interface afterwards from the contour line identified with the interface.

\begin{figure}
\centering { \includegraphics[scale=0.4]{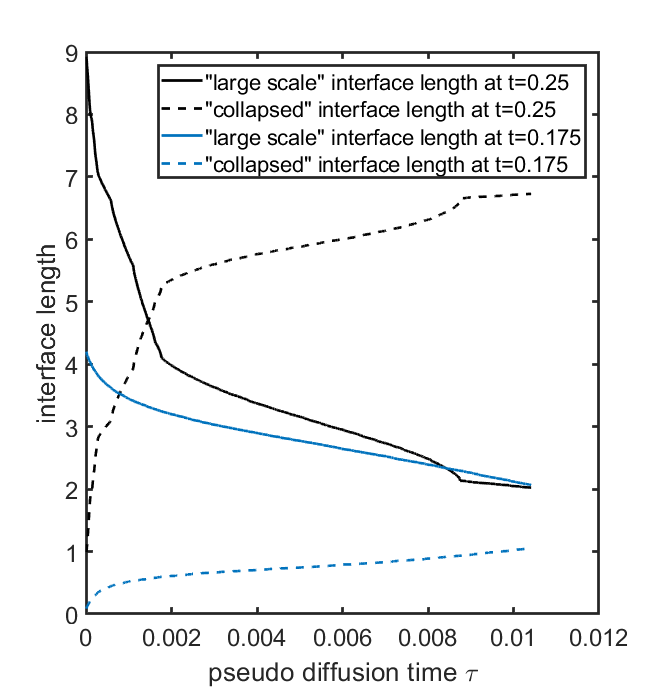} }
\caption{A plot of interface length versus  pseudo-diffusion-time. The solid line is the large scale interface length while the dashed line is the collapsed interface length.   }
\label{Figure-InterfaceLength}
\end{figure}

There are, of course, other possibilities to move the interface as the index function is smoothed. Instead of approximating the time derivative as done above, we can use equation (\ref{diffusion1}) and replace the time derivative by the Laplacian. Computing the Laplacian on the fixed grid and interpolating it to the interface  works and should, in principle, handle interacting interfaces correctly. We have, however, not found it as robust as the process described above.
Since the velocity of an interface in a diffusion field is  given by its curvature, we could  also move it  that way. It is, however, first  of all, generally difficult  to accurately evaluate the curvature and secondly, an interface evolved by its curvature is not guarantied to follow the appropriate contour as is the case with our approach, and numerical errors can therefore accumulate.

To show how the coarsening works, we apply it to the large amplitude stage of a  two-dimensional Rayleigh-Taylor instability. The computational domain is a  rectangle of dimensions $ 1 \times 2$ with periodic side boundaries and rigid top and bottom, and resolved by a regular $128 \times 256$ grid. The density and viscosity of the heavy and light fluids are $\rho_1 = 10$, $\mu_1  =0.02$, and $\rho_0=5$ and $\mu_0= 0.01$, respectively. Gravity acceleration is $g_y=-100$ and surface tension is $\sigma=0.25$. At the initial time the velocities are zero but the elevation of the interface is perturbed by $y(x)=1+0.1 \sin(2\pi x)+0.1 \sin(4 \pi x)$. 
 Figure \ref{Figure-Interface} shows the simplification of the interface at times $t=0.175$ and $t=0.25$. The filtering is shown as an evolution in pseudo time $\tau$, starting with the unfiltered interface on the left and progressing to the right. Stopping at a given pseudo-time correspond to a given filter size. The red line is the tracked interface and we also show  contours of the diffused index function. The gray traces are inactive (or ``collapsed'')  interfaces left over as the simplification progresses. 
 In the second frame, at $\tau=0.0004$, only the very smallest scales have been eliminated and the large scale structure of the interface is intact, except that at the later time the upward moving protrusion seen at the earlier time has just detached from the lower fluid. 
 In the third frame significantly more simplification has taken place and the downward protrusion moving downward at the earlier time has separated from the rest of the heavy fluid at the later time, so at the  later time the flows consists of a relatively flat interface and light fluid blob moving up and heavy blob moving down.
In the forth frame, at $\tau=0.0104$, the interface has been smoothed significantly at the earlier time and at the later time it consists of a nearly flat interface and one blob of the light fluid moving upward and a downward moving mixed region.  If we continued to smooth the interface, eventually we end up with a flat interface and mixed regions moving up and down.

The interface length is shown versus pseudo-time in figure \ref{Figure-InterfaceLength}, for both times shown in figure \ref{Figure-Interface}. The interface length decreases as the interface is simplified, as expected. In addition to the length of the interface separating regions of different phases (solid lines) we also plot the length of interfaces  that have ``collapsed'' and no longer serve as a phase boundaries (dashed lines). The length of those increases with time and the figure shows that the collapse takes place rapidly, leading to kinks in the curves.  The interface length is reduced in two ways: by  straightening the interface since diffusion corresponds  to interface motion by mean curvature and by interfaces disappearing as close by interfaces collapse onto each other. For this problem, the plot shows that most of the shortening of the interface is due to interfaces collapsing.

\begin{figure}
\centering { \includegraphics[scale=0.31]{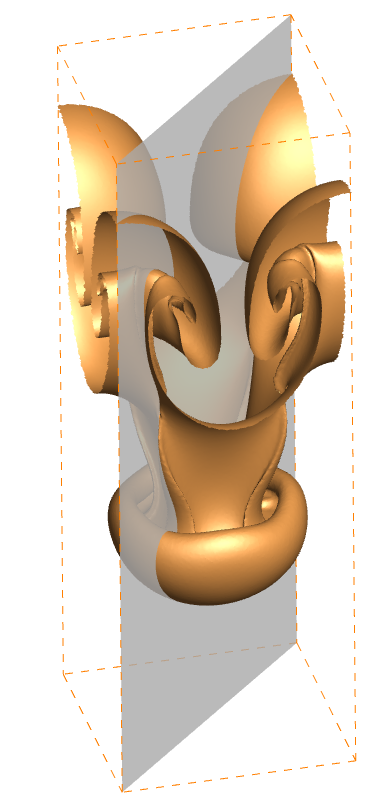}   \includegraphics[scale=0.26]{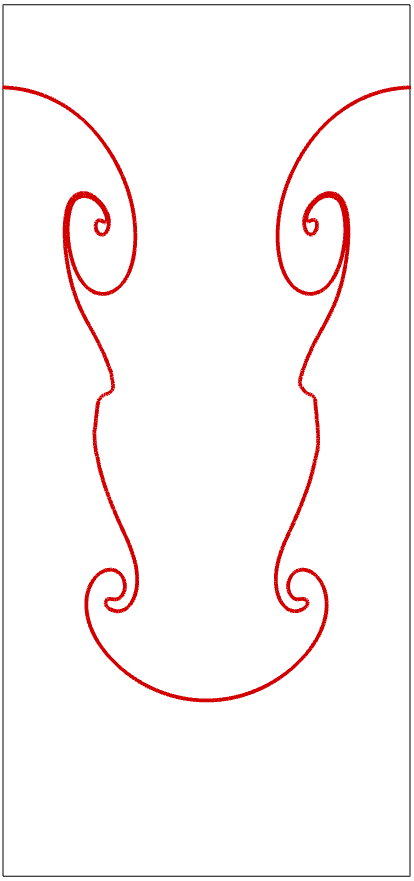}   \includegraphics[scale=0.287]{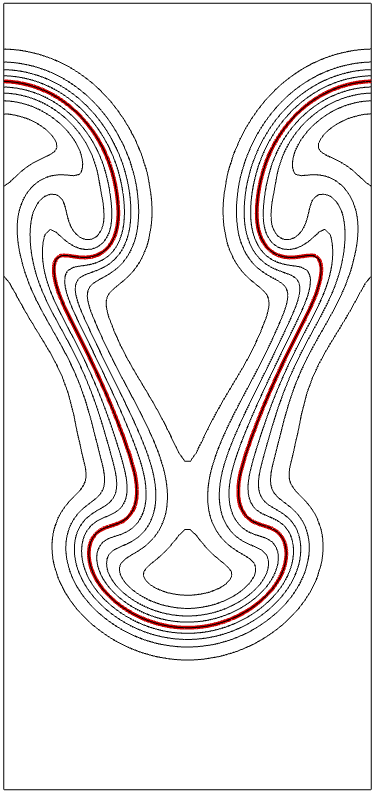}   \includegraphics[scale=0.305]{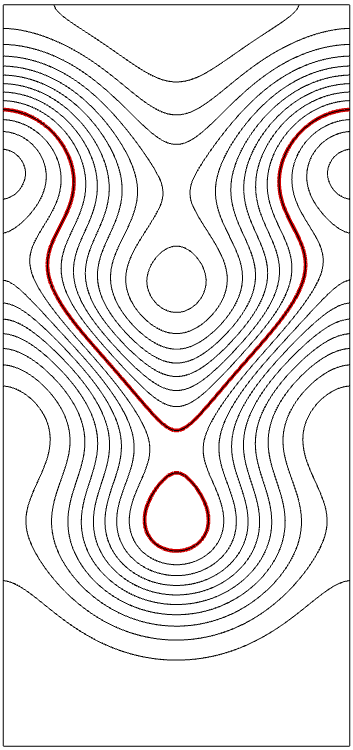}   \includegraphics[scale=0.287]{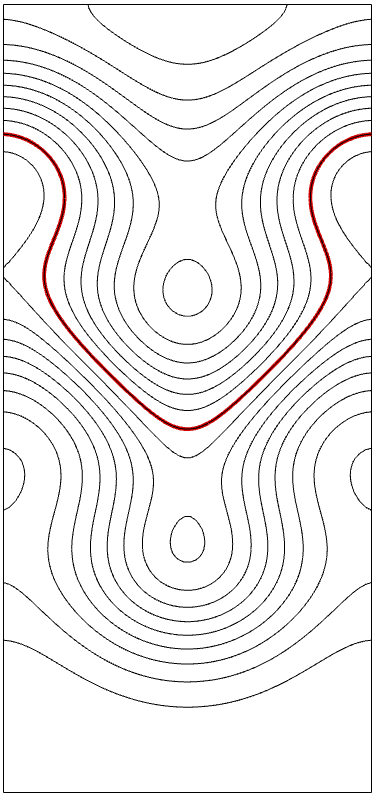} }
\caption{The  evolution of the interface  in pseudo-time as the interface is smoothed by diffusion for a three-dimensional Rayleigh-Taylor instability. The original interface is shown in the frame on the left and the contours for the index function in the gray plane are shown in the  four frames on the right for $\tau=0$ (original), $\tau=0.0026$ ($\Delta = 0.25$), $\tau=0.0204$ ($\Delta = 0.7$), and $\tau=0.0267$ ($\Delta = 0.8$). The red line is the intersection of the interface with the  plane.}
\label{Figure-3D-1}
\end{figure}

 Although we have chosen to show the simplification for two-dimensional flows, the process described above applies to fully three-dimensional flows as well and in figure  \ref{Figure-3D-1} we show the simplification for a Rayleigh-Taylor instability at one ``real'' time, versus pseudo-time. In the frame on the left the three-dimensional interface is shown along with a plane cutting diagonally through the computational domain. The interface and contours of the index functions in the gray plane are shown for three pseudo-times in the frames on the right. 
The computational domain is a periodic hexahedron of dimensions $ 1 \times 1 \times 3$ with periodic side boundaries and rigid top and bottom,  resolved by a regular $64 \times 64 \times 192$ grid. The density and viscosity of the heavy and light fluids are $\rho_1 = 10$, $\mu_1  =0.08$, and $\rho_0=5$ and $\mu_0= 0.04$, respectively. Gravity acceleration is $g_y=-100$ and surface tension is zero. At the initial time the velocities are zero but the elevation of the interface is perturbed by $z(x,y)=1.75+0.12 \times [ \cos(2\pi x)+ \cos(2 \pi y) ]$. The  small scales have been eliminated at  $\tau=0.0026$, at  $\tau=0.0204$ a downward moving drop has separated from the heavy fluid, and in the last frame ( $\tau=0.0267$) this drop has been mixed with the lighter fluid, as seen for the two-dimensional case.

\subsection{Smoothing the Flow Variables}
As the interface moves, field variables such as density and momentum are left behind on the opposite side of the interface. Variables that switch side are taken to belong to the small scales and we can account for their  evolution in many ways. Here we smooth those and ``blend'' them with the original field. Thus, we need to diffuse those variables, but only on one side of the interface. To do so we again solve a diffusion equation, but to prevent diffusion across the interface, we put the diffusion coefficients around the interface equal to zero in a thin band containing the interface. We denote the new diffusion coefficient by $\tilde D$ and since the diffusion coefficient is now not constant, we need to solve
\begin{equation}
\frac{\partial \phi }{ \partial  \tau} = \nabla \cdot \tilde D  \nabla \phi,
\end{equation}
where $\phi$ is the flow variable that we are smoothing. Smoothing the index function again, but using $\tilde D$ instead of a constant $D$, gives the volume fraction $\alpha$ of the phase where $\chi=1$ for the unfiltered field.

Unlike for single phase flow, where we generally filter the velocity field, here we filter the momentum, since it is the conserved quantity.  This can lead to nonzero divergence at the interface and to correct for that we recompute a pressure field needed to make the velocity divergence free. For  the mixture model used here, where both phases have the same velocity, the mixture velocity should be divergence free.  For a more advanced model, such as the drift flux model, we would have slip and the velocity used here becomes the  mass weighted velocity which is not necessarily divergence  free. However, once the slip is known, the volume source can be computed.

The complete smoothing process involves evolving the index function and the flow variables in pseudo-time $\tau$ for long enough to achieve the desired smoothing. In addition to diffusing the index function to move the interface (equations \ref{diffusion1} and \ref{interfacevelocity1}), we also evolve a copy of the index function to find the void fraction by using $\tilde D$, which consists of the constant diffusion coefficient  $D$ modified by putting it to zero at and around the interface. Similarly, we evolve the momentum using a pseudo-pressure field to enforce incompressibility. 

Denoting the filtered variables by a tilde, the equations to be evolved in pseudo time ($\tau \le T$) to generate a coarse flow field are:
\begin{align}
\label{line1} 
&  \tilde \chi (\tau =0) = \chi; \quad  \alpha (\tau=0) = \chi;  \quad D=1.0; \\
\label{line2} 
&\frac{\partial \tilde \chi }{ \partial \tau }  = \nabla^2 \tilde \chi ; \qquad {\bm u}_f= -\frac{1 }{ \Delta  \tau } \Biggl( \frac{ (\tilde \chi_I- \tilde \chi)}{ \vert \nabla \tilde \chi  \vert^2}
\Biggr) \nabla \tilde \chi ; \qquad \frac{d {\bm x}_f}{ d \tau} = {\bm u}_f  ;  \\
\label{line3} 
&\tilde D ({\bm x}, \tau) = D \hbox{ modified by setting } D({\bm x}_f )=0; \\
\label{line4} 
&\frac{\partial  \alpha }{ \partial \tau }  = \nabla \cdot \tilde D \nabla \alpha ;   \\
\label{line5} 
&\frac{\partial }{ \partial \tau } (  \widetilde{\rho {\bm u} } )   = -\nabla \tilde p + \nabla \cdot \tilde D \nabla ( \widetilde{\rho {\bm u} } ); \quad  \hbox{with} \quad \nabla \cdot \tilde{\bm u}=0.
\end{align}
We assume that $ \widetilde{\rho {\bm u} } = \tilde \rho \tilde {\bm u}$, that the density is given by $\tilde \rho = \alpha \rho_1 +(1-\alpha) \rho_o$, and that the filtered velocity is incompressible.  The second equation (\ref{line2}) is the smoothing of the index function described above. The third equation (\ref{line3}) is the modification of the diffusion coefficient by setting it to zero where the interface is.  In equation (\ref{line4}) we diffuse a copy of the index function to find the volume fraction, and the last equation (\ref{line5}) is the evolution of the momentum using a projection method to enforce incompressibility by finding the appropriate pressure.  Enforcing incompressibility of the coarse field would be difficult if the flow field is smoothed directly using a filter (\cite{Toutantetal2009b}) but seems both important  and  reasonable to do.
If we take the slip velocity to be zero, as in the simple mixture model used in the next section, there is only one velocity and the divergence is zero.
Once  the  interface has been moved by the smoothing, the coarse filtered index function $\tilde \chi$ identifying the large scale phase distribution is reconstructed by putting it equal to $1$ in one fluid and $0$ in the other.
When the coarsening leaves most of the phases separated by a sharp interface, we assume that relatively small amount of phase 0 is mixed with phase 1 and vice versa. For the coarse field the index function $\tilde \chi$ identifies the different fluids but $\alpha$ is equal to the diffused index function $\chi$. If there is no mixing, $\alpha =\tilde \chi=0$ in fluid $0$ and $\alpha =\tilde \chi=1$ in fluid $1$. 
We define a perturbation void fraction $\alpha' =\alpha - \tilde \chi$ which takes negative values in fluid 1 and a positive values in fluid 0 and quantifies how much of one fluid is mixed in  the other fluid. In general we expect $\alpha'$ to consists of isolated regions of positive and negative values and to be zero for most of the flow field. 

\begin{figure}
\centering { \includegraphics[scale=0.255]{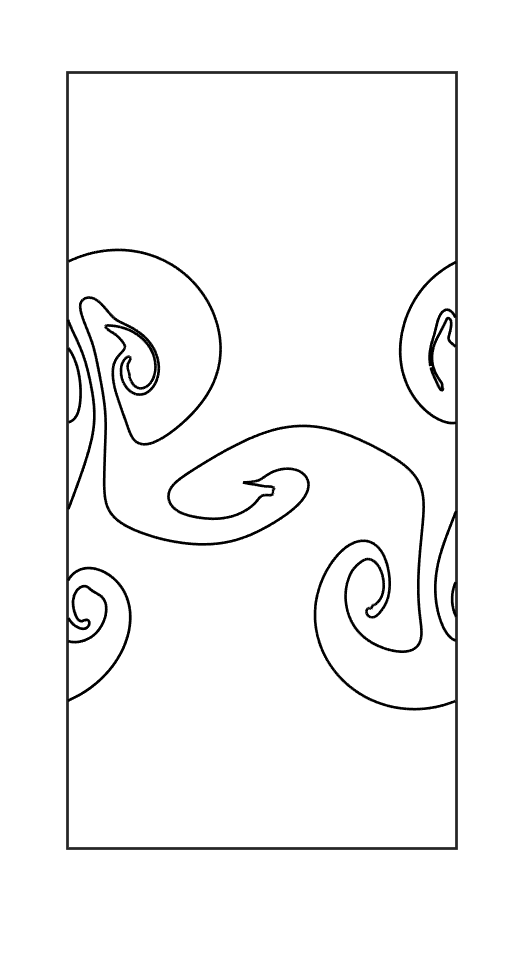}    \includegraphics[scale=0.268]{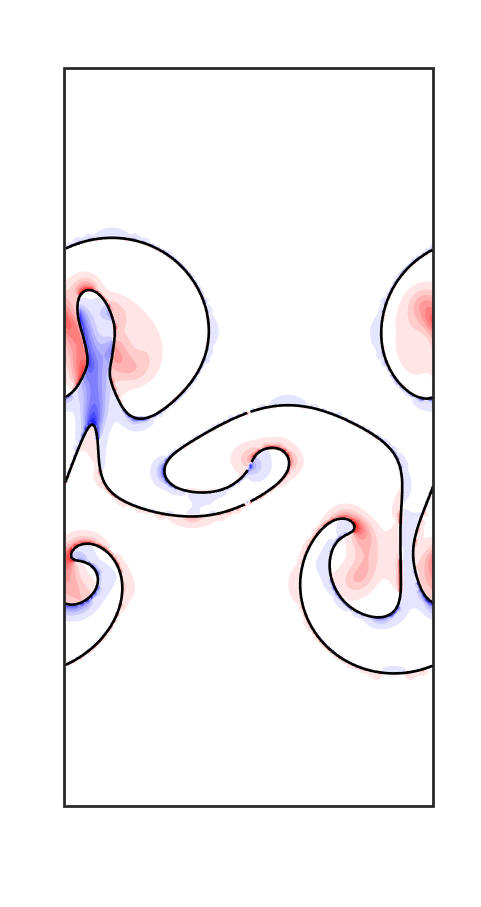}  \includegraphics[scale=0.268]{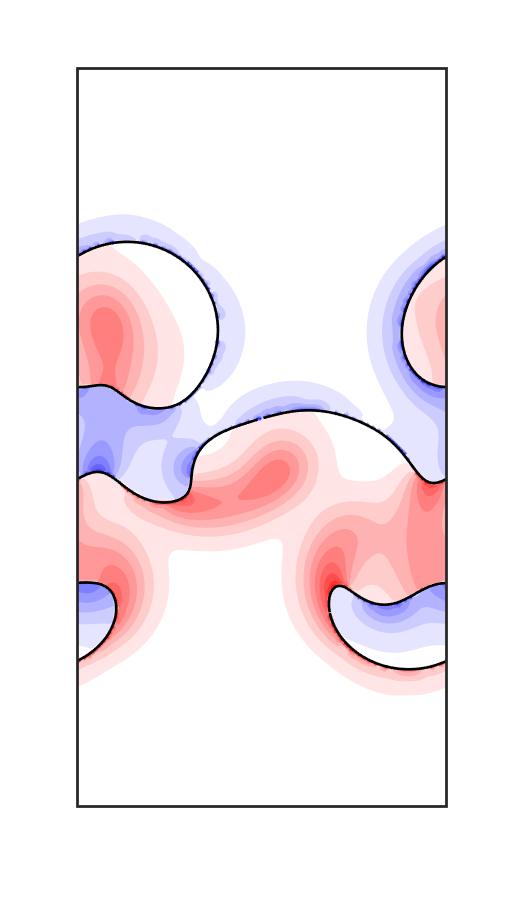}   \includegraphics[scale=0.268]{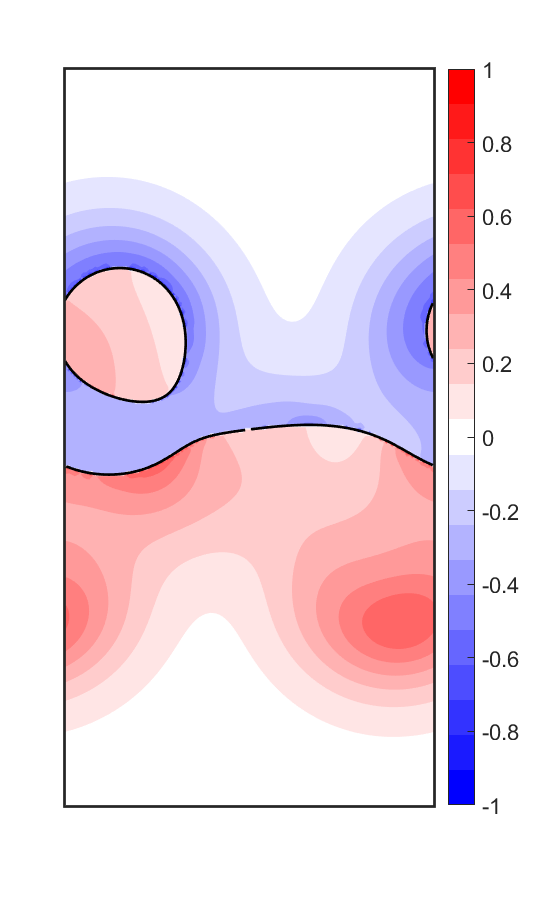}  }
\caption{The evolution of the perturbation volume fraction field ($\alpha'$) in pseudo-time by nonlinear diffusion, for the later time shown in figure \ref{Figure-Interface} ($t=0.25$) and the same values of $\tau$. }
\label{Figure-Voidfraction}
\end{figure}

Figure \ref{Figure-Voidfraction} shows the original sharp interface (left frame) and the perturbation volume fraction at three stages of smoothing, for the second time shown in  figure \ref{Figure-Interface} and at the same pseudo times. Where the interface is moved by the smoothing, one phase is ``left behind'' and is mixed with the phase originally on that side of the interface. Initially most of the mixing takes place near high curvature regions of the interface, where it is retreating rapidly, but as small scale features are eliminated, we see regions of high mixture fractions inside each domain.
We note that for interfaces that have collapsed into thin sheets, we do not put the diffusion to zero, although we have the option of doing so. 

\begin{figure}
\centering { \includegraphics[scale=0.255]{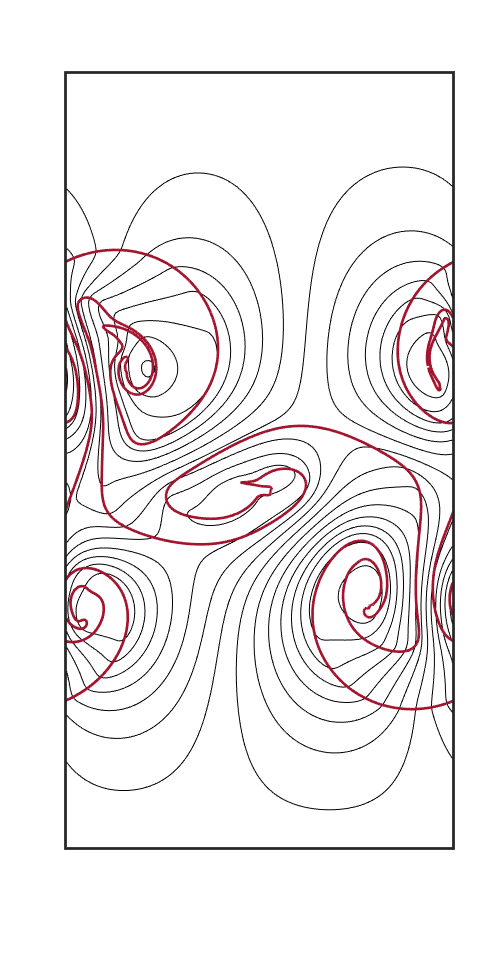}    \includegraphics[scale=0.255]{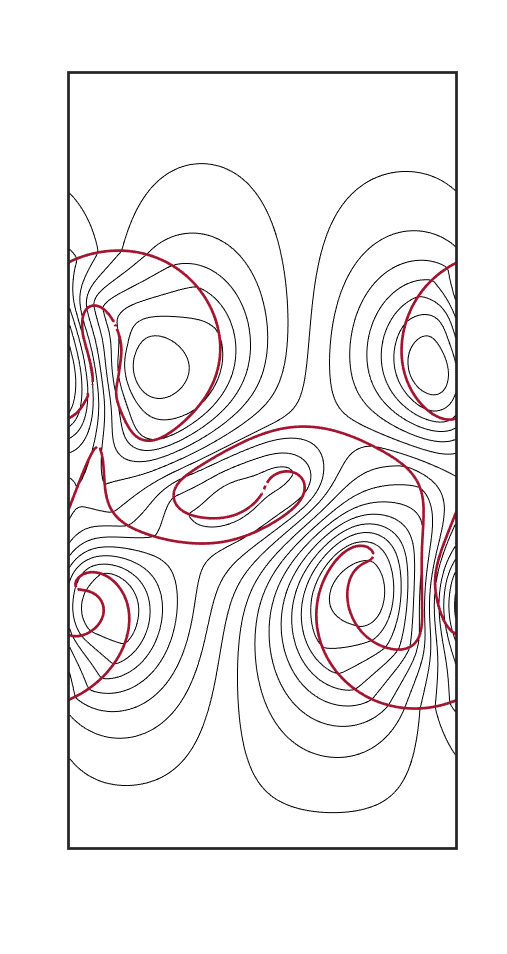}  \includegraphics[scale=0.255]{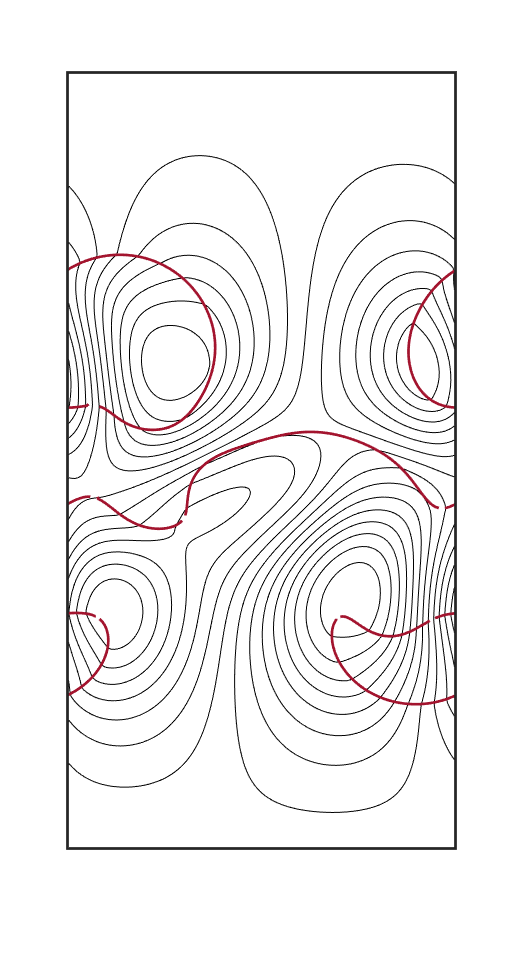}   \includegraphics[scale=0.255]{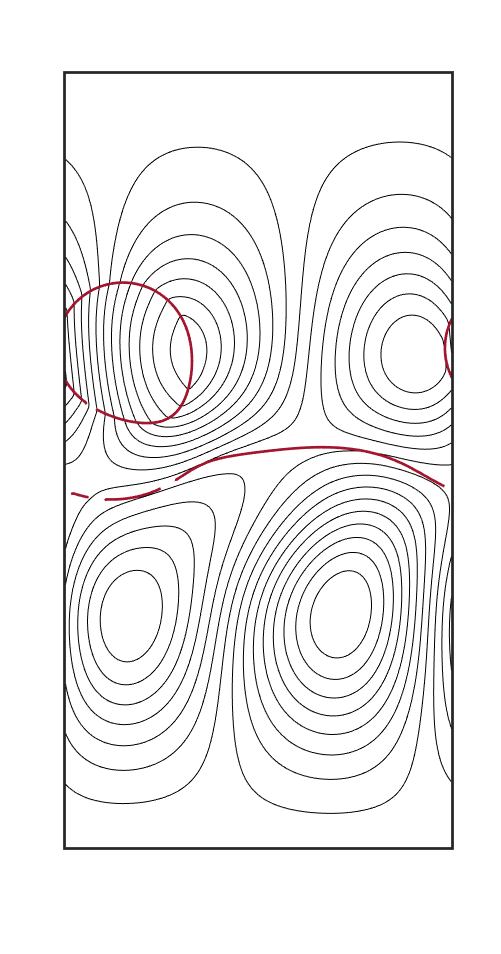}  }
\caption{The  evolution of the stream function in pseudo-time as the interface is smoothed by diffusion, for the later time shown in figure \ref{Figure-Interface} ($t=0.25$) and the same values of $\tau$. }
\label{Figure-Streamfunction}
\end{figure}

In figure  \ref{Figure-Streamfunction} the streamfunction computed from the velocity field is shown, first for the unfiltered velocity (left frame) and then for the smoothed velocity (frames 2-4)  at the same pseudo times as in figure \ref{Figure-Interface} and  \ref{Figure-Voidfraction}. Since we enforce incompressibility when we smooth the flow, the normal velocity remains continuous across the simplified interface. For modest smoothing, where only the smallest scales have been eliminated the flow field remains close to the original one (second and third frame) but for aggressive smoothing as in the last frame the flow field above and below the interface has been simplifies to a pair of counter rotating vortices, driven by the upward moving blob and the downward moving mixed region.

\begin{figure}
\centering { \includegraphics[scale=0.255]{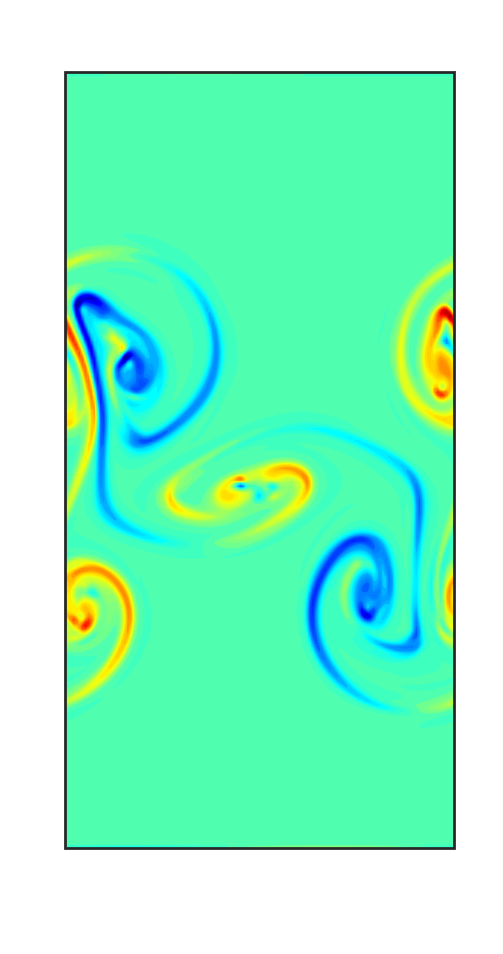}    \includegraphics[scale=0.255]{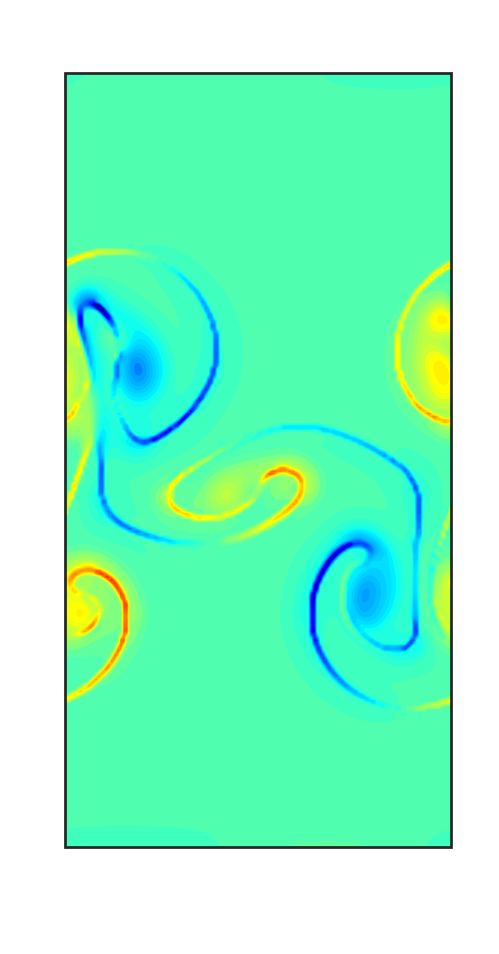}  \includegraphics[scale=0.255]{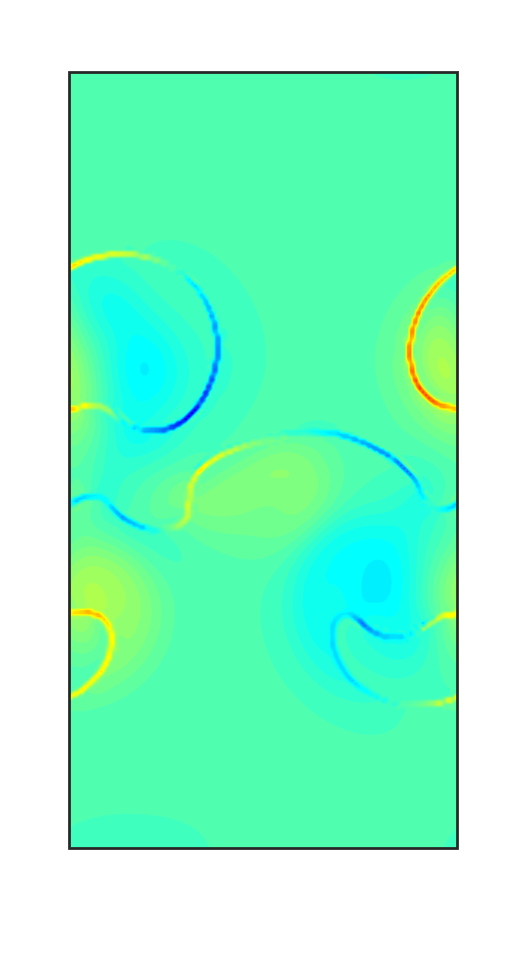}   \includegraphics[scale=0.255]{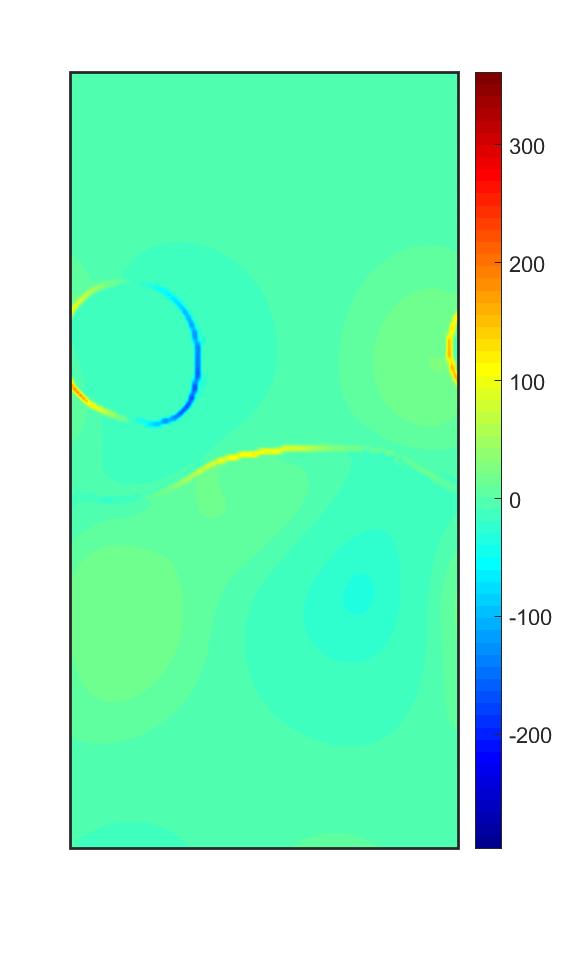}  }
\caption{The  evolution of the vorticity in pseudo-time as the interface is smoothed by diffusion, for the later time shown in figure \ref{Figure-Interface} ($t=0.25$) and the same values of $\tau$. }
\label{Figure-Vorticity}
\end{figure}

Figure \ref{Figure-Vorticity} shows the vorticity field computed from the velocity field, first for the original velocity (left frame) and then for the smoothed velocity (frames 2-4) at the same pseudo times as in figures \ref{Figure-Interface} -- \ref{Figure-Streamfunction}. Since the flow is driven by  baroclinic vorticity generated at the interface, the concentration is highest there and indeed, nearly all the original un-smoothed vorticity is at the interface. As small scale features are eliminated, the vorticity  is mixed with the bulk, although the concentration remains highest at the interface. For the most aggressive smoothing in the right frame, we see pairs of faint counter-rotating vortices in the bulk on either side of the interface, corresponding to the vortices seen in the plot of the streamfunction in figure  \ref{Figure-Streamfunction}. 

We can generate field values for other quantities than those show here. The interfacial area, for example,  is one important measure of the simplification of the phase distribution and we can construct an area field by differentiating the unfiltered index function. By filtering it we get the area concentration. Similarly, we can filter components of the area tensor. The volume fraction and the area concentration would allow us to obtain the  Sauter Mean diameter and equivalent number density, if we assume that the smoothed phase consists of spherical drops or bubbles.

\section{Evolving the Coarse Field} 

The main purpose of coarsening the flow field is to provide data for modeling the evolution of the coarse flow. Such models come in various forms and the purpose of the present paper is not to explore the many possibilities in any detail. We do, however, include one simple example in this section.

The homogeneous mixture model, where the phases are assumed to be completely mixed and move with the large scale fluid velocity is probably the simplest model imaginable. In our implementation a sharp interface is assumed to separate the large scales, and the small scales are represented by mixtures embedded in the large scale regions, so the conversion between large resolved scales and modeled small scales must be captured. The coarse interface moves with the coarse velocity when no conversion happens, but when transfer of the index function between large and small scales takes place, the interface and fluid velocity are different. We generally expect this to happen in high curvature regions. There is only one large scale velocity which is found by solving a momentum equation, supplemented by the incompressibility conditions:
\begin{equation}
\frac{\partial ( {\tilde \rho \tilde {\bm u} } ) }{ \partial t }   +\nabla \cdot (\tilde \rho  \tilde {\bm  u}  \tilde {\bm  u} ) = -\nabla \tilde p + \tilde \rho {\bm g} + \nabla \cdot {\bm \tau}_e \quad  \hbox{ and } \quad \nabla \cdot \tilde  {\bm u} = 0.
\label{model2}
\end{equation}
Here we have lump into ${\bm \tau}_e$  all transport terms due to viscous stresses, surface tension and small scale velocity fluctuations. 
The density field everywhere is given by
\begin{equation}
\tilde \rho = (\alpha' + \tilde \chi ) \rho_1 +(1-\alpha' - \tilde \chi) \rho_o,
\end{equation}
where $\alpha'$ is the perturbation volume fraction introduced earlier and is evolved by an advection-diffusion equation:
\begin{equation}
\frac{\partial  \alpha' }{ \partial t } +  {\bm  u}   \cdot \nabla \alpha' =  \nabla \cdot \tilde D_e \nabla \alpha' +S_I.
\label{model3}
\end{equation}
$\tilde D_e$ is an effective diffusion term, adjusted to match the spreading of the mixed region in the coarsened field and modified to prevent diffusion across the interface by setting $\tilde D_e ({\bm x}_I)=0$.  $\tilde D_e$ can depend on $\alpha$, $\tilde \chi$, and other flow variables, but  here we will take it to be a constant.
The interface source term is related to the relative speed of the interface by
\begin{equation} 
S_I=[\chi] ({\bm u}_f - {\bm u}_I ) \cdot {\bm n}=[ \chi] \Delta u_I
\end{equation}
where ${\bm u}_f$ is  the fluid velocity and ${\bm u}_I$ is  the interface velocity, and $ \Delta u_I =({\bm u}_f - {\bm u}_I ) \cdot {\bm n}$. 
The interface moves by ${d {\bm x}_I / dt} = {\bm  u}_f - \Delta u_I {\bm n} $ so $ \Delta u_I $ determines the ``production'' of small scales where the interface does not exactly follow the fluid velocity.

To close the equations for the coarse field, we must develop relationships that describe how $\Delta u_I$, $D_e$ and ${\bm \tau}_e$ depend on the coarse variables. In general, we expect that those would be found by comparing the evolution of the solution for the coarser field with the evolution of the coarsened fully resolved solution and correlated by machine learning, for example. Here, however, we have simply guess a value for $\tilde D_e$, and taken the stress term in the momentum equation to be the same as for the unfiltered flow, including the surface tension. Finding $\Delta u_I$ requires a more complex approach. We expect most of the small scale production to take place in high curvature regions and since
motion by mean curvature is equivalent to diffusion, as noted earlier, the simplest approach is to use the same approach as we used for the original coarsening and solve a diffusion equation in pseudo time, after the solution has been advanced assuming the interface moves with the fluid velocity. Once the interface has been moved, the phase ``left behind'' becomes inside the other phase becomes small scale flow.

\begin{figure}
\centering { \includegraphics[scale=0.252]{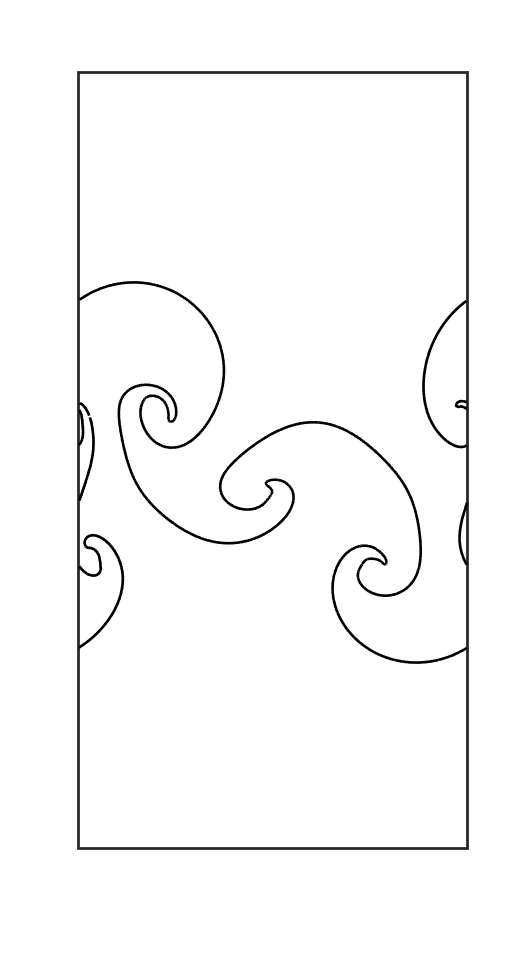}    \includegraphics[scale=0.265]{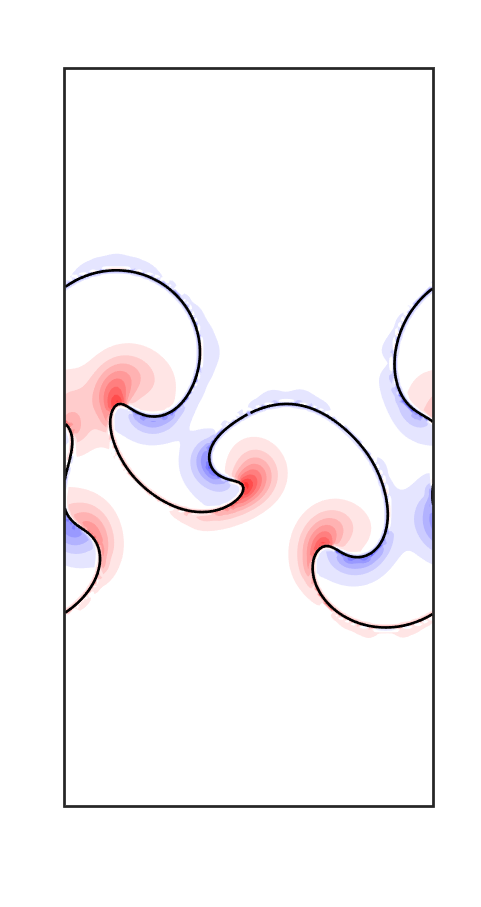}  \includegraphics[scale=0.265]{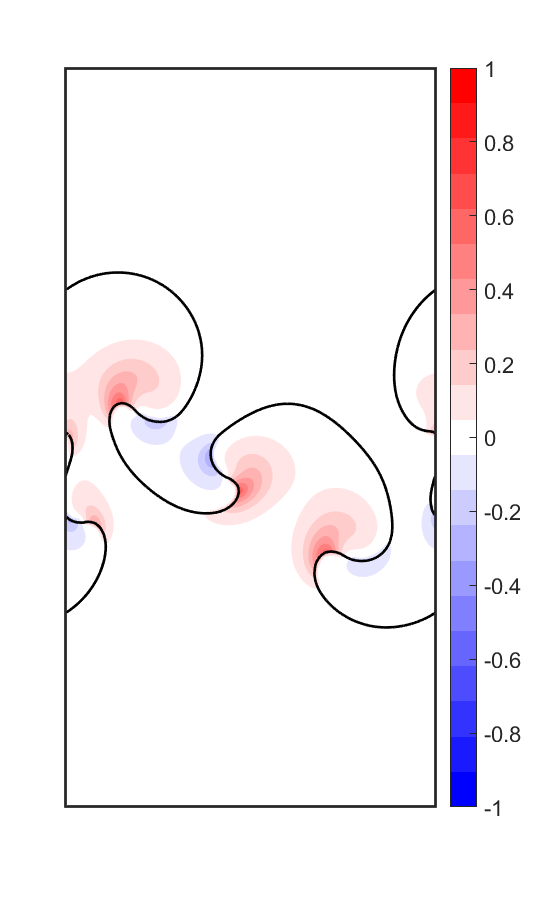}  }
\caption{The  Rayleigh-Taylor instability at time $t=0.2$. The left frame shows the unfiltered interface, the middle frame shows the interface and the perturbation volume fraction obtained by smoothing the flow by diffusion. The solution in the right frame is obtained by the simple mixture model described in the text. }
\label{Figure-2D-RT}
\end{figure}

Figure \ref{Figure-2D-RT} shows a large amplitude stage of two fluids undergoing a Rayleigh Taylor instability at time $t=0.2$. The first frame shows the original unfiltered interface. the middle frame shows the interface and the perturbation volume fraction obtained by smoothing the flow by diffusion using $\Delta = 0.18$, and the right frame is the solution evolved from the same initial conditions using the simple homogeneous mixture model with $\tilde D_e=0.05$ and $\Delta u_I$ found by diffusion in pseudo time, every twenty time steps, with the modification that we leave the interface untouched if it moves less than $0.15 \times h$, where $h$ is the grid spacing. Here, $\Delta t = 6.1 \times 10^{-5}$ and $\Delta \tau= 6.1 \times 10^{-6}$ and we take ten steps in pseudo time. The model equations are solved by the same method as used for the fully original  flow and on the same grid. The frequency of the interface modification was selected such that the interface matches roughly with the filtered interface, but no effort was made to match the effective diffusivity closely. At earlier times the agreement is also good, but after the time shown here the filtered interface starts to undergo topology changes and a more sophisticated selection of the model parameters would be necessary.

\section{Conclusion} 
We discuss a strategy to coarsen multiphase flows in a consistent way while retaining a sharp, but simplified interface. The motivation is the possibility of developing closure modeling through machine learning by working directly with the coarsened field where an explicit form relating the closure terms to the fully resolved flow is not needed. Regions without any mixing can be treated using standard turbulence models and  mixed zones, where one phase is embedded in the other, can be treated in a variety of ways, such as assuming a homogeneous mixture (as done here), by a drift flux model, an Eularian two fluid model, or representing the embedded phase as Lagrangian particles.

We note that here we represent the interface by connected marker points that are moved as the smoothing progresses. To  simplify the interface we could also simply smooth the index function and then reconstruct the interface from the contour of the average value. For the  other steps, where we separate the large and small flow scales by diffusion and we need to evolve the interface along with the field values,  having the interface makes this relatively straightforward.  This could, however, presumably also be done in the absence of explicit tracking of the interface by reconstructing the interface from the average contour at every step in pseudo time.

Although the motivation for the coarsening strategy presented here is reduced order modeling of the flow, it also provides us with a tool that should be useful in analyzing the distribution of scales in multiphase flows. As we coarsen the flow, we separate it into small and large scales, which can be examined separately as the coarsening is varied. Other possibilities include following a different contour than the average one to ``skeletonize'' the flow. We hope to explore some of those possibilities in later studies.

\vskip 10pt

{\bf Acknowledgement:} 
This research was supported in part by NSF grant CBET-1953082.

\bibliography{GrandBib,TryggvasonBib}

\end{document}